# How search engine marketing influences user knowledge gain

Knowledge gain and search engine marketing

Development and empirical testing of an information search behavior model

Sebastian Schultheiß

Hamburg University of Applied Sciences, sebastian.schultheiss@haw-hamburg.de

People use search engines to find answers to questions related to their health, finances, or other socially relevant issues. However, most users are unaware that search results are considerably influenced by search engine marketing (SEM). SEM measures are driven by commercial, political, or other motives. Due to these motivations, two questions arise: What information quality is mediated through SEM? And how is collecting documents of different quality affecting user knowledge gain? Both questions are not considered by existing models of information behavior. Hence, the doctoral research project described in this paper aims to develop and empirically test an information search behavior model on the influences of SEM on user knowledge gain and thereby contribute to the search as learning body of research.

**CCS CONCEPTS** •Information systems~World Wide Web~Web searching and information discovery~Web search engines

## 1 INTRODUCTION

Search engines are integral to everyday life and influence user decision-making on socially relevant issues, for example, health-related [5] or political topics [4]. This influence is not shaped by search engine providers alone since several other stakeholder groups impact what is shown on the search engine result pages (SERPs), namely users, content producers, and search engine marketing [17:80]. Search engine marketing (SEM) covers search engine optimization (SEO) and paid search marketing (PSM), both acting hand in hand with content producers. Search engine optimization (SEO) is "the practice of optimizing web pages in a way that improves their ranking in the organic search results." Paid search marketing (PSM), on the other hand, "is operated by search engines in the form of sponsored or paid results" [15:3110].

Thus, SEO and PSM pursue the same objective – making online content visible to search engine users through optimized organic or paid results [1]. SEO and PSM directly address the behavior pattern of users, as users strongly focus on what is immediately visible on the SERP [e.g., 3,9] when selecting a result. Although SEM directly influences search engine result pages, many users are unaware of these measures. This is particularly true for SEO, as a representative survey of German Internet users shows. While around 68% of users are aware of Google's advertising business model, only 43% know that results can be influenced beyond the placement of ads, i.e., through SEO measures [14].

SEO and PSM professionals closely interact with content producers who are in need of attention to their content – whether for commercial, political, or other reasons [17:81]. Due to the variety of intentions underlying the use of SEM measures, the question arises to what extent these intentions influence the information quality of optimized (SEO) or paid (PSM) results. As noted by van der Sluis et al. [21], information quality (IQ) is often defined rather vaguely in the literature and characterized based on several criteria, such as usefulness or accuracy [e.g., 16]. To assess the criteria, people use indicators, i.e., observable attributes within a given document that serve as clues as to whether or not a given criterion is met [22]. The criteria can be grouped into those concerning the intrinsic quality, i.e., accuracy, objectivity, believability, reputation, and those relating to the representation of the information, i.e., interpretability, ease of understanding, concise representation, consistent representation [12:43].

To summarize, people use search engines to answer essential questions. People usually examine only a few prominently placed results on the SERP when searching, with the results being subject to considerable influences from SEM. SEM measures are driven by commercial, political, or other motives, which suggest an influence on the documents' quality and thus on user knowledge gain, i.e., the difference between the knowledge before and after a search task [7]. These influences are not yet addressed in models of information behavior. However, such a model is important to gain an understanding of (1) how information quality is mediated through SEM and (2) how collecting documents of different quality affects user knowledge gain. This doctoral research project aims to develop and empirically test an information search behavior model on the influence of SEM on knowledge gain and to contribute to the search as learning body of research [10].

## 2 RESEARCH QUESTIONS AND HYPOTHESES

Figure 1 illustrates the components of my research project and how they are interrelated. Below the figure, the research questions (RQs) and the corresponding hypotheses are explained.

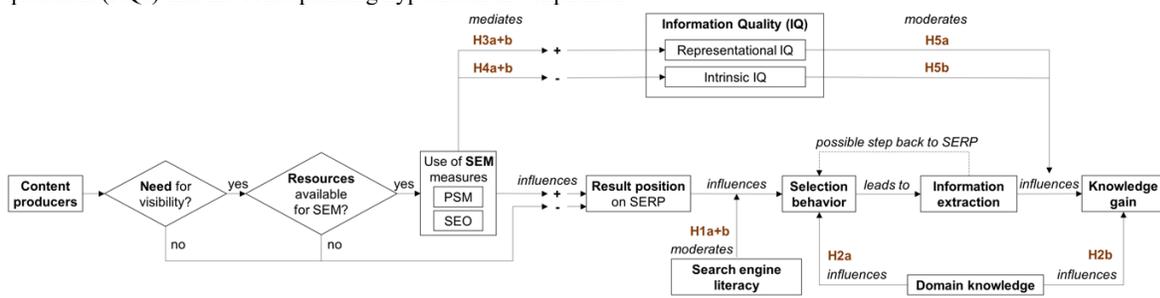

Figure 1: Hypotheses within the context of the doctoral research project

**RQ1:** *How does search engine literacy influence user selection behavior?* As indicated by previous work [19], user with low awareness of PSM are more likely to select ads. Furthermore, users with higher SEO awareness are expected to be guided less strongly by the position of an (organic) result and thus consider more organic result ranks (e.g., results below the fold).

- H1a: The higher the PSM awareness, the fewer ads are selected.
- H1b: The higher the SEO awareness, the wider the range of selected organic results.

**RQ2**: *How does domain knowledge influence user knowledge gain?* Users with domain knowledge are expected to select relevant documents more efficiently [2] or select no documents at all when already knowing the answer to the task. Domain knowledge is also likely to reduce knowledge gain for most subjects, as noted by, e.g., Gadiraju et al. [7].

- H2a: The greater the domain knowledge, the fewer results are selected.
- H2b: The greater the domain knowledge, the lower the knowledge gain.

**RQ3:** *From the use of SEM measures, what conclusions can be drawn on the representational information quality of the content?* Due to quality criteria for ads [8] and SEO measures [20], optimized representations are to be expected for documents performing PSM *or* SEO measures.

- H3a: Results that originate from an advertisement have a higher representational information quality than results that do not perform PSM or SEO measures.
- H3b: Results that perform SEO measures have a higher representational information quality than results that do not perform PSM or SEO measures.

**RQ4:** *From the use of SEM measures, what conclusions can be drawn on the intrinsic information quality of the content?* Conducting PSM or SEO measures is driven by the motivation to attract the internet users' attention, for example, to a specific medication. Consequently, a more one-sided representation of information and, thus, a lower intrinsic information quality is assumed.

- H4a: Results that originate from an advertisement have a lower intrinsic information quality than results that do not perform PSM or SEO measures.
- H4b: Results that perform SEO measures have a lower intrinsic information quality than results that do not perform PSM or SEO measures.

**RQ5:** *How does selected documents' representational and intrinsic quality influence user knowledge gain?* Both representational and intrinsic information quality of the selected documents will influence user knowledge gain.

- H5a: Selected documents with higher representational information quality are accompanied by a higher knowledge gain than documents with lower representational information quality.
- H5b: Selected documents with higher intrinsic information quality are accompanied by a higher knowledge gain than documents with lower intrinsic information quality.

**RQ6:** *How do the investigated variables (SEM presence on SERP, information quality of the documents, search engine literacy, and domain knowledge) affect user knowledge gain?* This research question focuses on the interaction effects of the variables. Interesting evaluations include how knowledge gain is affected by documents with *opposing* qualities (e.g., low intrinsic but high representational quality).

## 3 METHODS

To answer the RQs, an online experiment combined with an expert evaluation will be conducted. The aim of the experiment is to examine SEM measures and knowledge gain for causal relationships. Before, during, and after the experiment, factors that might influence the relationship are collected, including selection behavior and search engine literacy. As described in RQ3 and RQ4, an influence of SEM measures on the information quality of the documents is assumed. Therefore, the quality of all results included in the experiment will be evaluated. Figure 2 shows the preparation plan for both methods.

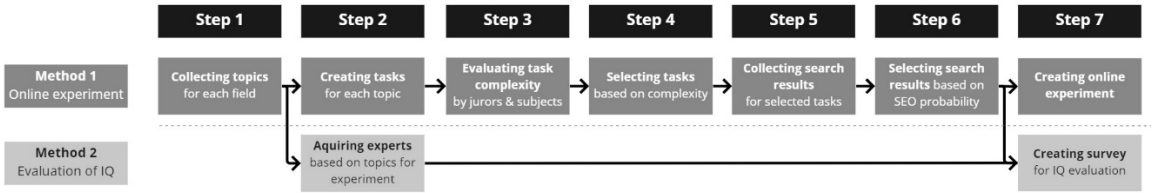

Figure 2: Preparation plan for both methods

### 3.1 Preparing the online experiment

Next, the steps of the preparation plan for the online experiment (method 1) illustrated in Figure 2 are explained.

1. *Collecting topics*: For the socially relevant fields of health, politics, environment, and consumer protection, $N = 3$ topics are selected for each field. To find topics, lists of terms will be used, such as lists of diseases provided by public authorities or lists of environmental topics provided by environmental associations. This will result in topics such as asthma or climate change.
2. *Creating tasks*: For all topics, topical aspects from Google's "people also ask" box are collected, such as "What helps with asthma at night?" These topical aspects are then used to construct tasks that prompt subjects to find information on the particular topical aspect. A high number of tasks ($N = 100$ or more each topic) is intended to ensure that the tasks cover a wide range of complexity levels.
3. *Evaluating task complexity*: Since task complexity heavily influences user behavior [e.g., 18], first, jurors will be invited to evaluate the complexity of all tasks. Second, to test the reliability of the judgments, tasks that are potentially suitable for the experiment are tested in a small laboratory study with subjects.
4. *Selecting tasks*: Based on the complexity judgments by the jurors and subjects, tasks of different complexity levels are selected, e.g., one complex and one simple task each topic.
5. *Collecting search results*: For all selected tasks, a pool of organic and paid search results is collected.
6. *Selecting search results for online experiment*: From the pool of collected search results, organic and paid search results are selected for the experiment according to their SEO probability as classified by the SEO classification tool developed in our research group[1]. A rule-based classifier determines the probability of SEO for a specific web page based on SEO indicators. An example of an SEO indicator is the use of analytics tools, which can be identified from the HTML code of a web page.
7. C*reating the online experiment*: In creating the experiment, three aspects are of particular importance, namely (1) creating manually curated SERPs, (2) measuring knowledge gain, and (3) creating pre- and post-experiment surveys.
(1) For each task, SERPs will be manually curated using the results that have been selected in the previous step. Several SERP conditions are built to investigate the effect of SEM measures on knowledge gain. The conditions will differ in the result types presented (organic results only, organic results and ads) and in the optimization probability (e.g., high, medium, low) of search results.
(2) In the literature, a variety of methods are identified for measuring knowledge gain, such as multiple-choice tests or essays [6:104]. Since the online experiment will be conducted with a large sample (see 3.3), qualitative methods that require the analysis of written summaries or similar are hardly feasible. Therefore, a quiz based approach with pre-task and post-task knowledge tests seems to be most appropriate [e.g., 11].

---

[1] https://github.com/searchstudies/seoeffekt

(3) Pre- and post-experiment surveys for measuring domain knowledge, search engine literacy, and demographics will be developed. For the search engine literacy survey, preliminary work can be followed up [e.g., 19].

### 3.2 Preparing the evaluation of information quality

The preparation of the information quality evaluation is directly related to steps one and six of the preparation of the online experiment. Once the topics of the online experiment are specified (step one), the preparation phase of the information quality evaluation begins by acquiring experts from the respective fields. Once the search results have been selected for the online experiment (step six), the survey for information quality assessment can be set up. It is intended to reuse existing questionnaire items, e.g., AIM quality (AIMQ) by Lee et al. [13].

### 3.3 Procedure of the online experiment

For the online experiment, a sample representative of the German online population with a sample size of between $N = 1,000$ and 2,000 people will be used, as we did in previous studies [e.g., 14] in cooperation with market research companies. The experiment has the following procedure:

1. Pre-experiment survey: demographics, domain knowledge
2. Experiment: Subjects will be asked to complete tasks of different complexity while their selection behavior is captured. The tasks are performed in three steps. First, before beginning a task, a pre-task survey is conducted to determine the subject's knowledge. Then the task together with the curated SERP is shown to the subject. After the task has been completed, the post-task survey for measuring knowledge gain is carried out.
3. Post-experiment survey: search engine literacy (conducted at the end to not influence user behavior in the experiment)

### 3.4 Procedure of the evaluation of the information quality

Using an online questionnaire, experts will receive screenshots of web pages and items to assess their quality. The questionnaire will also contain items to characterize the experts' field of expertise, e.g., by asking for their current position and years of experience.

## 4 PROGRESS AND FUTURE PLANS

The concept of the PhD project has been defined and the literature research will soon be completed. Pre CHIIR 2023, I plan to have preliminarily formulated the literature sections. I will also further differentiate the methods preparation plan to address individual aspects during CHIIR specifically. Post CHIIR 2023, the methods according to the plan outlined above will be prepared so that the feedback received at the conference can be incorporated. The main emphasis of the preparation is on developing the tasks and the pre- and post-task knowledge tests, the search engine literacy survey, and the questionnaire for the information quality evaluation. Simultaneously, the subjects and experts for both methods are acquired, and approval by the local ethics committee is obtained.

## ACKNOWLEDGMENTS

I would like to thank my supervisors Prof. Dr. Dirk Lewandowski and Prof. Dr. Vivien Petras for their ongoing support. This work is funded by the German Research Foundation (DFG Deutsche Forschungsgemeinschaft), grant number 467027676.